# "She was useful, but a bit too optimistic": Augmenting Design with Interactive Virtual Personas

Paluck Deep[1], Monica Bharadhidasan[1], A. Baki Kocaballi[1]

[1]School of Computer Science, University of Technology Sydney, Australia

**Abstract**

Personas have been widely used to understand and communicate user needs in human-centred design. Despite their utility, they may fail to meet the demands of iterative workflows due to their static nature, limited engagement, and inability to adapt to evolving design needs. Recent advances in large language models (LLMs) pave the way for more engaging and adaptive approaches to user representation. This paper introduces Interactive Virtual Personas (IVPs): multimodal, LLM-driven, conversational user simulations that designers can interview, brainstorm with, and gather feedback from in real time via voice interface. We conducted a qualitative study with eight professional UX designers, employing an IVP named "Alice" across three design activities: user research, ideation, and prototype evaluation. Our findings demonstrate the potential of IVPs to expedite information gathering, inspire design solutions, and provide rapid user-like feedback. However, designers raised concerns about biases, over-optimism, the challenge of ensuring authenticity without real stakeholder input, and the inability of the IVP to fully replicate the nuances of human interaction. Our participants emphasized that IVPs should be viewed as a complement to, not a replacement for, real user engagement. We discuss strategies for prompt engineering, human-in-the-loop integration, and ethical considerations for effective and responsible IVP use in design. Finally, our work contributes to the growing body of research on generative AI in the design process by providing insights into UX designers' experiences of LLM-powered interactive personas.

## 1. INTRODUCTION

*"I think there are a lot of subtlety or details, things that cannot be conveyed from the description... I don't think you can really think or act like [the persona]" —Design practitioner quoted in* (Matthews et al., 2012).

This statement captures a long-standing limitation in static persona-based design: the disconnect between well-crafted descriptions and the lived complexity they attempt to represent. While personas remain a staple in human-centred design (HCD), their static, document-based form often fails to support the evolving needs of iterative, collaborative workflows. They cannot be questioned, probed, or adapted in real time. When novel ideas emerge or ambiguities arise, static personas offer little recourse. In response, this paper investigates the potential of Interactive Virtual Personas (IVPs): Large Language Model (LLM)-powered, conversational agents that simulate user perspectives through real-time dialogue. We examine their potential to extend the utility of personas by enabling dynamic, situated interaction in contemporary design workflows.

Personas offers hypothetical yet representative user profiles including that persona's goals, needs, desires and motivations to guide design teams in maintaining focus and empathizing with their target audience (Cooper, 1999; Preece et al., 2015). When well-grounded in research and effectively communicated, personas can shape product strategies, inform feature prioritizations, and bolster team consensus (Adlin & Pruitt, 2010). Personas have been widely used in reference to design (Dharwada, 2006), design and development of consumer healthcare technologies (LeRouge et al., 2013), online data analytical systems (Salminen et al., 2018), audit management system for aircrafts (Dharwada et al., 2007), security systems (Faily et al., 2021) and AI education for college and university students (Fujita et al., 2023).

The use of personas to guide design decisions can increase the productivity and profitability of an organization (Drego et al., 2010). Numerous surveys across diverse organizations have demonstrated that persona development and utilization are widely adopted practices, seamlessly integrated into the design processes supporting the creation of various products (Nielsen et al., 2015). However, personas relying on static text and visuals may fall short in representing the dynamic nuances of real users' goals and behaviours. They are limited by their one-way, impersonal nature, making it challenging for designers to engage in an ongoing dialogue that might surface new needs, confirm ambiguous assumptions, or explore multiple design directions (Matthews et al., 2012). This one-way, non-interactive format restricts designer engagement primarily to passive reading. Designers cannot spontaneously probe deeper into a persona's stated needs or motivations, explore hypothetical edge-case scenarios that arise during ideation, or seek clarification on ambiguities encountered when evaluating prototypes (Zhou et al., 2024), making it challenging to support deep, ongoing empathy and understanding beyond the initial familiarization phase. These limitations can lead to slowed design iteration due to the difficulty in obtaining quick, persona-grounded feedback. They may also reduce designer engagement with personas that feel inert rather than like active representations of the target audience. This gap highlights a critical need for user representation methods that offer greater dynamism, interactivity, and adaptability to support the fluid, iterative nature of HCD.

In response to these shortcomings, recent work has begun exploring interactive personas: AI-powered agents that allow users to “talk to the persona,” introducing a dynamic, bidirectional interaction model (Gu et al., 2025; Kaate et al., 2025; Sabbaghan & Brown, 2024; Sun et al., 2024; Zhou et al., 2024). Salminen et al. (2021), in their 15-year survey of persona research, identify human–persona interaction as one of the major emerging research directions, centred on the key question: “What benefits do persona users gain from talking to the persona?” This shift towards interaction reflects growing recognition that dialogic engagement may support greater empathy, uncover hidden user needs, and enable more contextualized and iterative design processes. Recent advances in LLMs present an opportunity to transform these static personas into interactive virtual agents that can converse with design teams in real time. Researchers have shown the potential of LLMs to emulate various user roles, suggest design ideas, and generate context-rich content for diverse application areas (Hong et al., 2023; Kocaballi, 2023; Gu et al., 2025). Yet, few studies (Gu et al., 2025; Sun et al., 2024) have explored whether an LLM-powered interactive persona can address some of the core limitations of conventional static personas including the lack of spontaneity, limited empathy-building, and insufficient granularity of user feedback across multiple stages of HCD process. While prior work has demonstrated the potential of LLM-powered interactive personas in supporting discrete design activities such as ideation and user simulation (Gu et al., 2025; Sabbaghan & Brown, 2024; Sun et al., 2024; Zhou et al., 2024) and their limitations such as hallucinations (Kaate et al., 2025), these approaches often remain limited in their ability to maintain continuity and contextual adaptability across the multiple phases of HCD processes. A critical gap remains: the absence of empirical studies examining how LLM-powered interactive virtual personas that support continuity, adaptability, and naturalistic voice interactions function across multiple stages of HCD. This gap limits our understanding of how such personas can augment real-world design practices.

To address this gap, in this work, we present an LLM-powered (GPT-4) interactive persona referred to as interactive virtual persona (IVP) that i) can engage users through its voice-enabled conversational interface; and ii) maintain a continuous interaction with user across multiple design activities. The IVP simulates a fictitious user, "Alice," an owner of a sustainable farm-to-table restaurant, whom UX designers can interview, brainstorm with, and gather feedback from throughout the design cycle. The IVP is integrated across key stages of the design process including user research, ideation, and prototype evaluation, enabling us to examine how the IVP adapts to diverse design activities and content domains, as well as how designers actively manage their relationship with the persona through conversational prompts. Furthermore, its voice-enabled, multimodal interface is intended to support more naturalistic interactions that mirror real-world design practices, offering new opportunities for naturally occurring dialogic engagement with interactive personas. Through semi-structured interviews with eight UX professionals after their interactions with the IVP, we address the following research questions (RQs):

**RQ1**: *How do LLM-powered IVPs support or hinder UX designers' practices?*

**RQ2***: What are LLM-powered IVPs' perceived benefits and limitations?*

Our findings reveal that IVPs enhance efficiency in early design stages by providing rapid, human-like feedback. However, designers raised concerns about biases, over-optimism, and the difficulty of maintaining authenticity without real stakeholder involvement. This work contributes empirical insights into the evolving role of generative AI (GenAI) in the HCD, offering actionable guidelines for integrating IVPs as complementary tools in design practice.

## 2. BACKGROUND

### 2.1 Personas and their limitations

Personas have been extensively researched and applied across various domains of design and development, demonstrating their utility in diverse contexts beyond digital products. Studies have showcased their value in automotive design (Marshall et al., 2015), consumer healthcare technology (LeRouge et al., 2013), and assistive technology design (Subrahmaniyan et al., 2018). Several studies have explored the role of empathy maps and user personas in design: Rohmiyati et al. (2023) used empathy maps from interviews and questionnaires to design user-centric e-resource systems for college libraries; Goswami & Goswami (2022) employed six user research techniques to curate personas and enhance app usability; and Friess (2012) analysed the impact of personas on design decision-making through an ethnographic case study. These studies illustrate the broad applicability of personas as tools for informing design decisions and supporting a user-centric approach within design teams.

While personas have been applied across various domains, their core utility and challenges are most evident within design and development processes. Studies illustrate how personas are used as tools for informing design decisions and supporting a user-centric approach within design teams. They are recognized for their potential in HCD (Salminen et al., 2022) such as early stage consolidation of user information (Anvari & Richards, 2015), brainstorming (Dow et al., 2006), prototyping (Bødker et al., 2012), and stakeholder communication (Jensen et al., 2017). Personas can distil complex user research into relatable user models that facilitate effective team communication and construction of shared mental models (Holtzblatt et al., 2004), support emphatic immersion (Miaskiewicz & Kozar, 2011), lead to better design decisions (Pruitt & Adlin, 2005), and challenge organisational assumptions and guide strategic prioritization of design efforts (Miaskiewicz et al., 2008). However, they present several notable drawbacks throughout their lifecycle. Salminen et al. (2018) provided a succinct summary of the limitations of personas in their creation, evaluation and use, including high development costs, biases of creators, reliance on non-representative data, lack of credibility, inconsistent and irrelevant information, and frequent under-utilisation. During creation, they can be time-consuming and expensive to develop, may reflect the biases of their creators, and sometimes rely on nonrepresentative data (Vincent & Blandford, 2014). Once developed, personas often face scepticism regarding their credibility and accuracy and may include information that is irrelevant or inconsistent for decision-making (Chapman & Milham, 2006). Even when personas are rigorously created and evaluated, they risk being underutilized in practice (Rönkkö et al., 2004) or harnessed for organizational politics and to reinforce preconceived notions (Hill et al., 2017).

Even after being rigorously created and evaluated, personas risk being underutilized in practice (Rönkkö et al., 2004) or harnessed for organizational politics and to reinforce preconceived notions (Hill et al., 2017). For example, despite significant investment in persona development, Friess's ethnographic study found that designers rarely invoked personas during decision-making (Friess, 2012). Use of personas was largely confined to designated scenario walkthroughs and driven mainly by the designers who created them. Similarly, Rönkkö et al. (2004) provide detailed accounts of extensive persona projects that, despite considerable investments of resources, were ultimately never implemented or meaningfully integrated into design workflows. One critical factor of underutilization can be associated with personas' static nature limiting their utility in dynamic, iterative workflows. For instance, Matthews et al. (2012) found that designers often discard personas during ideation due to the fact that personas are often seen as "abstract" and "impersonal," failing to provide the critical detail and empathy needed for effective design, with designers preferring "first-hand experience with users" over abstractions. Their static nature restricts interaction, making it challenging for designers to deeply engage with the persona's perspective beyond initial reading. The abstract and impersonal format of textual descriptions can hinder the development of genuine empathy, potentially leading to a superficial understanding of user needs. Furthermore, static personas, often presented as lengthy documents, can suffer from information overload, diluting focus on critical design elements and hindering efficient application in fast-paced design cycles (Matthews et al., 2012).

Many of these challenges especially underutilisation, abandonment and superficial implementation can potentially be mitigated by more dynamic and engaging user representation methods that can overcome the inherent constraints of static personas and facilitate a deeper, more actionable understanding of user needs throughout the HCD process. Our work aims to address these limitations by exploring interactive, conversational, LLM-powered personas, aiming to inject dynamism and facilitate a more empathetic engagement with user representations in design.

### 2.2 AI-generated and AI-simulated Personas

While developing personas manually through qualitative methods have been the primary persona creation method (Brickey et al., 2012), more automated, algorithmic and data-driven persona methods (DDPD) have been proposed as a response to the limitations of manual persona development (Salminen et al., 2021). DDPDs rely on machine learning techniques such as k-means and hierarchical clustering to segment users based on large-scale behavioural demographic data. They provide several advantages such as enhanced objectivity, decreased cost, updatability, and scalability (Salminen et al., 2021). Despite their benefits, prior research has indicated several limitations including data quality and availability (Salminen et al., 2021), bias and exclusion (Salminen, Jung, et al., 2020), and lack of depth and nuance (Holden et al., 2017). More recently, with the rapid advancements in GenAI and LLMs, a new line of GenAI based persona creation methods has emerged (e.g., (Holzinger et al., 2022; Salminen et al., 2023)). These AI-generated personas can reduce the manual effort needed to enrich the persona descriptions (Kaate et al., 2025), improve the representativeness of personas (Sun et al., 2024), support expression of personality traits (Jiang et al., 2024) and facilitate human-persona interaction (Shin et al., 2024).

Within this new line of research, a distinct branch of research emerged: AI-simulated personas (Gu et al., 2025; Kaate et al., 2025; Sun et al., 2024; Zhou et al., 2024). These are personas that are simulated by GenAI technologies. AI-simulated personas may or may not involve AI-generated personas. A manually developed static persona description can be simulated by GenAI systems. The main motivations of simulating a persona stem from the need to improve the sense of realism (Zhou et al., 2024), enrich representativeness (Jiang et al., 2024; Zhou et al., 2024), support persona-engagement (Kaate et al., 2025) and increase empathy towards persona (Gu et al., 2025) through a naturalistic conversational interaction between persona users and personas. The AI-simulated persona research has been primarily motivated by the advancements in LLM-based chatbots with more flexible and human-like conversational capabilities (Prpa et al., 2024). LLMs' ability to generate contextually relevant and coherent text, answer questions, and engage in dialogue opens up new possibilities for human-computer interaction, including applications within the HCD process. Studies have explored the use of LLMs for mental wellbeing chatbots (Kumar et al., 2022), demonstrating their capacity for thoughtful, non-judgmental, and easily understandable interactions. Prompt engineering has emerged as a key technique for tailoring LLM chatbot behaviour and capabilities (Kumar et al., 2022; Zamfirescu-Pereira et al., 2023), allowing for customization of chatbot identity, intent, and behaviour. Table 1 summarises the key distinctions between AI-generated and AI-simulated personas. In this study, we employ both approaches: first generating, and then simulating, the persona "Alice," with the full process detailed in Sections 3.2 and 3.3.

**Table 1.** Distinguishing AI-Generated from AI-Simulated Personas

| Dimensions | AI-generated personas | AI-simulated personas |
|---|---|---|
| Core Function | Creating or enriching persona descriptions. | Enabling real-time, conversational interaction with a persona. |
| Primary Artifact | Static profiles with goals, needs, pain points, quotes, demographics and images. | Interactive AI agent producing dialogic behaviour involving responses and role-play. |
| Input | Raw user data (e.g., analytics, surveys) or high-level prompts describing a target user group. | A complete, pre-existing persona description (which can be manually created or AI-generated). |
| Output | A static document or artifact describing a user persona. | A dynamic, interactive dialogue where the AI agent speaks "as" the persona. |
| Example from this paper | LLM-generated persona description for “Alice” (Appendix A.1a). | LLM-simulation of “Alice” that designers converse with (Sections 4.1 and 4.2). |

However, alongside their potential, LLM-powered chatbots also inherit limitations that are relevant when considering their application as interactive personas. Bias embedded in training data is a significant concern, potentially leading to stereotypical or skewed representations of users (Cheng, Durmus, et al., 2023; Ferrara, 2023; Sattele & Ortiz, 2024), causing designers to inadvertently reinforce existing prejudices or neglect diverse user perspectives. This poses a significant risk of reinforcing existing inequities rather than surfacing diverse user needs. Maintaining context over extended conversations and ensuring response consistency can also be challenging (Jung et al., 2025; Kocaballi, 2023), leading to persona interactions that feel fragmented, inconsistent, or generic, thereby diminishing their perceived realism and usability for informing nuanced design decisions. Furthermore, LLMs can exhibit hallucinations or inaccuracies in their responses (Amin et al., 2023; Jung et al., 2025; Zhou et al., 2024), potentially providing users with misleading insights or reducing trustworthiness of the AI-simulated persona (Kaate et al., 2025). For instance, a persona might convincingly describe emotional responses or life circumstances that are entirely invented by the model, despite being perceived as authentic. Perhaps most critically, LLMs, while capable of mimicking human conversation, lack genuine empathy and lived experience, raising ethical considerations about using them to represent and understand human users in the design process (Sorin et al., 2024). Reliance on such incomplete representations could result in products that fail to resonate with real users, highlighting the necessity of careful evaluation, verification, and supplementation of AI-generated insights. These potential pitfalls highlight the importance of carefully evaluating and mitigating the risks associated with using LLM-powered chatbots as interactive personas in HCD, ensuring responsible and ethical application of this technology.

### 2.3 Generative AI in design processes: Positioning Interactive Virtual Personas

Recent research established foundational insights into GenAI's role in design processes. Schmidt (2023) examined GenAI and LLMs' role in accelerating various aspects of interactive system development. Hong et al. (2023) explored how LLMs create consumer profiles for product design, while Kocaballi (2023) demonstrated ChatGPT's versatility in HCD tasks assuming roles of a designer, user and product. Specialized applications are also emerging, including conversational companions for older adults (Alessa & Al-Khalifa, 2023), personas for individuals with complex needs (Sun et al., 2024), and tools for generating persona narratives from interview data (Paoli, 2023) or streamlining user research workflows (Schuller et al., 2024; Shin et al., 2024). In training contexts, LLM-based personas are being explored for interview practice (Sabbaghan & Brown, 2024) and patient simulation in mental health training (R. Wang et al., 2024).

Our work on IVPs builds upon this growing body of research by specifically focusing on interactive and multi-modal LLM-powered personas as a direct augmentation to traditional static personas within the broader HCD process. Unlike previous work primarily focused on persona generation or specific design tasks, we investigate the end-to-end application of IVPs across multiple design activities, including user research, ideation, and usability evaluation. Furthermore, our emphasis on multi-modal interaction aims to enhance the sense of presence and naturalness in designer-persona interactions. While Zhou et al. (2024) developed Vivid Personas for real-time video interaction, our approach focuses on leveraging readily available text and voice modalities within a conversational framework, making IVPs more accessible and easily integrable into existing design workflows.

## 3. METHOD

Given the emerging and under-theorised nature of AI-simulated interactive virtual personas, this study adopted an exploratory qualitative approach (Miles et al., 2020). Such an approach is particularly well-suited to investigating novel design phenomena where the aim is to generate in-depth insights into behaviours, practices, and interpretative meanings, rather than to test predefined hypotheses (Braun & Clarke, 2021). IVPs combine three under-studied elements including interactive persona simulation, multimodal voice interaction and cross-phase deployment, so there is a lack of prior theory specifies which behaviours, benefits or breakdowns to expect. In such nascent areas, exploratory qualitative work is recommended because it enables flexible, inductive theorising grounded in participants' situated practices. Engaging professional UX designers as expert participants allowed us to draw on their situated, practice-based expertise to examine how IVPs are appropriated across distinct stages of the design process, including user research, ideation, and prototype evaluation. Semi-structured interviews supported the elicitation of experiential reflections and critical engagement with the IVP, enabling participants to articulate domain-specific concerns, conversational prompt-engineering strategies, and ethical judgments in their own terms. These qualitative insights, grounded in real-world usage scenarios, offer a rich empirical basis for conceptualising IVP integration and inform the development of future constructs and measures for subsequent quantitative or mixed-methods investigations. The study was approved by the Low-Risk Ethics Committee of University of Technology Sydney.

### 3.1 Participants

We used purposive sampling and recruited eight professional UX, UI, and product designers through targeted LinkedIn outreach and public calls in design-focused groups. Eligible participants had to be actively working in user experience roles, with a balance sought between junior (<5 years) and senior (>5 years) experience levels. This variation was intentional: junior designers could bring recent exposure to AI tools and emerging workflows, while senior designers could contribute deep domain knowledge and a critical lens shaped by pre-AI practices. Our goal was not statistical generalization but to capture a rich diversity of perspectives on the use of IVP. While the sample size is small, it reflects the qualitative, exploratory nature of the study and was sufficient to identify recurring patterns across distinct phases of the design process. We acknowledge the limitations of this approach, including potential self-selection bias and geographic concentration, which may limit broader applicability. Full participant details are presented in Table 2.

We recruited 8 UX designers (5 women, 3 men), based in Sydney, Australia (n=6) and Delhi and Noida, India (n=2). This distribution reflects the locations where our research team had the most established industry contacts. Geographic location was not a variable of interest in the study design. Their industry backgrounds spanned Fintech,

Edtech, Telecom, and IT, representing a diverse range of design contexts. Half of the participants had less than five years of experience, while the other half had more than five. For clarity, we refer to participants as P1J–P4J for Junior designers, and P5S–P8S for Senior designers. All participants reported prior exposure to at least one AI-based tool (e.g., ChatGPT, Gemini, Claude, Notion AI, Midjourney, Rabbit R1) in their professional and personal lives, showcasing a general familiarity with AI technologies. A detailed view of participant demographics and the AI tools they use is presented in Table 2. Before beginning the study, each participant was introduced to the IVP and completed a brief trial to familiarize themselves with its interface and voice/text responses.

**Table 2.** Participants' demographics information and AI tools they use

| Participant ID | Age | Gender | Occupation | Years of experience | Industry | Location | AI Tools Used |
|---|---|---|---|---|---|---|---|
| P1J | 29 | Female | Software tester and UI designer | 1 | IT | Sydney | AI in Bing, Canva AI, ChatGPT, Copilot |
| P2J | 24 | Female | Freelance UX designer and graphic designer | 2 | IT | Sydney | Adobe Firefly, Claude AI, Leonardo AI, Nightshade |
| P3J | 24 | Male | Founding designer | 4 | IT | Delhi | Chatsonic, Copilot, Gemini, Grammarly, Leonardo AI, Midjourney, Notion AI |
| P4J | 26 | Female | Full time UI/UX designer | 4 | Edtech | Noida | ChatGPT, Gemini, Midjourney |
| P5S | 33 | Female | Senior Experience Designer | 5 | Fintech | Sydney | DALL-E |
| P6S | 29 | Male | Full time UI/UX designer | 5 | Edtech | Sydney | Copilot, Midjourney, Notion AI, Otter AI, Photoshop AI, Twitter AI |
| P7S | 28 | Male | Freelance UI/UX designer | 6 | Fintech | Sydney | AI in Meta, Google Bard, Midjourney, Photoshop AI, Rabbit R1, Sora AI |
| P8S | 32 | Female | UX designer | 10 | Telecom and IT | Sydney | AI in Figma, AI in LinkedIn, AI in Miro |

### 3.2 Procedure

We developed an IVP prototype using a customised multimodal ChatGPT-4 (OpenAI, 2025), built with the GPTs editor introduced by OpenAI (*Introducing GPTs*, 2024). We configured the model with tailored instructions and behaviour settings to simulate a consistent persona. We used system messages, which are special initial prompts that guide the model's behaviour, to define the persona "Alice" with her background information (Appendix 1.a). There were also instructions emphasising human-like conversation, favouring concise and natural responses rather than the detailed, point-by-point answers typically produced by GPT models (Appendix 2). To anchor the study, we chose a food business scenario focused on sustainable dining, a domain that participants could readily envision and discuss without requiring specialized technical knowledge. The steps to create the final IVP were as follows:

**Initial persona draft**: We prompted ChatGPT-4 to produce a persona named Alice Rivera, including demographics, daily routines, motivations, pain points, and technology usage (Appendix A.1a). We used standard persona templates (Nielsen et al., 2015) and criteria for persona perception (Salminen et al., 2020) to inform the prompt design. The custom multimodal GPT, named Alice, integrates voice, text, and image capabilities, with voice serving as the primary mode of interaction. Users primarily engage with Alice through its voice user interface. It also supports image input, which is used for receiving feedback on user interface designs, and allows for text input and output when voice interaction is insufficient or unavailable.

**Pre-study pilot and iterative refinements**: We performed a brief pilot test to determine the right level of detail in persona descriptions to simulate the persona in the subsequent designer-persona sessions. Specifically, we posed a broad range of user-research questions (N=24) covering personal background, business motivations, operational processes, staffing, customer service, marketing, technology use, sustainability practices, and future goals (Appendix A.3) to both a detailed persona (Appendix A.1a) and a concise persona (Appendix A.1b) and qualitatively inspected their answers for naturalness and topical relevance. Although the performance was broadly comparable: the detailed persona offered slightly deeper domain-specific reasoning (e.g., partnering with local farmers to manage organic costs), and displayed a more reflective, "thinking-aloud" tone, whereas the concise persona produced smoother, higher-level summaries. Response length varied inconsistently and did not appear to align directly with persona description length. While the differences were limited, we selected the detailed version for the main study because it demonstrated greater potential for nuanced, domain-relevant reasoning. It is important to note this pilot was not intended as a rigorous evaluation of how persona description length influences response quality and authenticity; its sole purpose was to choose a suitable starting point for our designer-persona sessions.

Future work could systematically test granularity effects, but our study instead foregrounds the opportunities and concerns that emerge when UX designers engage with an LLM-simulated interactive persona across diverse design activities. Finally, to support more natural conversational behaviour, we refined system prompts to encourage human-like expressions, incorporate filler words (e.g., "umm," "ahh"), maintain continuity across multi-turn conversations, and generate shorter, more fluid responses without formal list structures for explanations (Appendix A.2).

**Final IVP Configuration**: The final instructions instructed GPT-4 to respond in the first person, avoid extraneous disclaimers, provide concise responses, and simulate thinking pauses (Appendix A.2). We then deployed this prototype on a mobile interface that allowed participants to communicate through both text and voice, thereby simulating conversations with a more lifelike user.

### 3.3 Data collection

We conducted 70–80 minutes-long semi-structured, one-on-one interviews with professional UX designers, divided into three key parts:

i) **Orientation and Practice (5–10 minutes):** We introduced the study and the IVP by providing a brief overview of its voice-based capabilities. Participants then practiced interacting with the IVP to familiarize themselves with its capabilities and limitations.

ii) **Task-Based Interaction (30–40 minutes):** Participants engaged with the IVP across three activities including user research, ideation, and prototype evaluation. centred on the persona's hypothetical business (a sustainable food service). They posed questions, uploaded the images of wireframe designs as part of prototype evaluation activity (Appendix A.4), brainstormed ideas, with all interactions audio-recorded and text transcripts captured.

iii) **Post-Task Interview (25–30 minutes):** Participants answered our interview questions, exploring participants' overall experience, perceived usefulness, impact on design activities, limitations, and suggestions for improving the use of live virtual personas in the design process. They also completed a 7-item Likert-scale survey evaluating factors such as authenticity, response diversity, suggestion feasibility, and usefulness across three activities (Appendix A.5).

### 3.4 Data analysis

We transcribed all interviews using Otter.ai (*Otter.ai*, 2025) and checked for accuracy. Using NVivo 14 software (NVivo, 2025), we analysed the interview transcripts using a reflexive thematic analysis (RTA) approach (Braun & Clarke, 2012) involving dataset familiarization, data coding, initial theme generation, theme development and review, theme refining and defining, and report write up. Two researchers independently engaged with the first four transcripts, immersing themselves in the data through repeated reading and generating inductive, data-driven initial codes that captured meaningful patterns in participants' experiences with IVP; no a priori codebook was applied. They then engaged in a reflexive discussion, exploring different interpretations and refining the coding framework in response to their evolving understanding of the data. Unlike consensus-seeking thematic analysis, which aims to minimize subjectivity through coder agreement, reflexive thematic analysis embraces the researcher's active role in meaning-making, allowing codes and themes to develop fluidly throughout the analytic process. This approach was particularly well suited to our study, which aimed to surface nuanced, situated insights into how professionals appropriate a novel, interactive LLM-driven persona. The flexibility of RTA supported the evolving, exploratory nature of our small dataset and allowed us to remain responsive to emergent meanings in participants' accounts. The lead researcher proceeded with coding the remaining transcripts, maintaining an iterative and flexible approach, where codes were adapted and refined as new insights emerged. After all transcripts were coded, the codes sharing similar conceptual meanings were grouped to form candidate themes. These candidate themes were then refined through ongoing reflexive discussions within the research team. Themes were retained when they demonstrated internal coherence, clear boundaries, and representation across multiple participants; others were merged, split, or discarded. For instance, the final theme of Over-optimism was developed from the initial, more literal code "Everything is feasible according to Interactive Persona." Similarly, the Collaborative design partner theme emerged not as a single code but as a synthesis of two concepts, including "Interactive Persona as design aid" and "Contextual interaction with designer." Throughout the process, ongoing reflexive engagement and discussions among the research team helped to deepen the interpretation of the data.

The quantitative survey data, collected immediately post-interaction, was intended to provide a structured view of participants' immediate perceptions of the IVP's characteristics and usefulness across the design tasks. This data serves to complement and contextualize the in-depth qualitative insights gathered from the semi-structured interviews. Basic descriptive statistics were calculated to summarize these quantitative responses. Given the exploratory nature of this qualitative study and the small sample size (N=8), inferential statistical analyses were deemed inappropriate. Such tests would lack sufficient statistical power, potentially leading to misleading conclusions. Therefore, the analysis focuses on descriptive summaries of the survey data to support the primary qualitative findings.

## 4. RESULTS

### 4.1 Survey results

Participants completed a 7-item Likert-scale questionnaire (1 = strongly disagree, 7 = strongly agree) after each of the three IVP tasks (User Research, Ideation, Prototype Evaluation), yielding 24 ratings per dimension (8 participants × 3 activities). Figures 1a and 1b visualise these data: Figure 1a shows median and IQR by activity; Figure 1b collapses across activities to compare junior (< 5 yrs) and senior (≥ 5 yrs) designers.

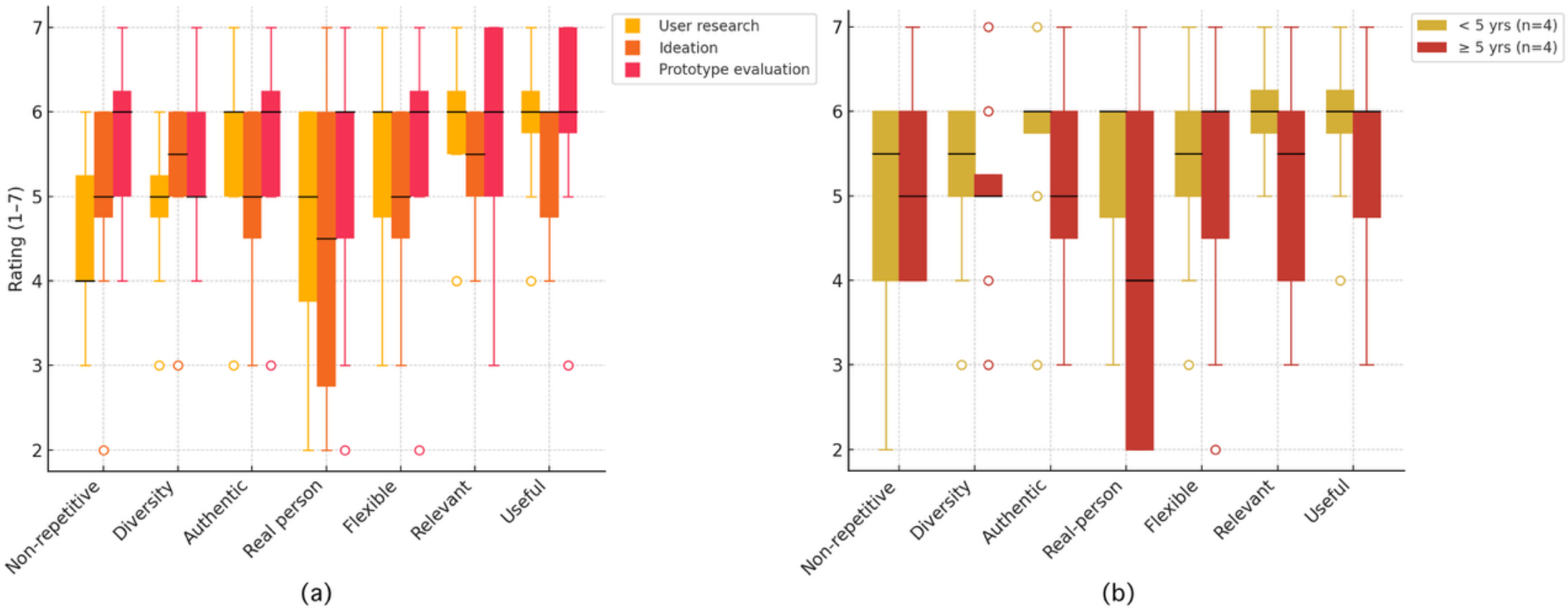

**Figure 1**. a) IVP survey ratings across design activities; and b) Junior and Senior designer perceptions of IVP. **Survey Items:** 1) *Non-repetitive:* 'The responses are not repetitive'; 2) *Diversity:* 'There is diversity in the responses'; 3) *Authentic*: 'The responses seem authentic to me'; 4) *Real-person:* 'I believe Alice is a real person'; 5) *Flexible:* 'The responses are flexible to my prompts'; 6) *Relevant:* 'The responses are relevant to my design work'; 7) *Useful:* 'I found the customised GPT to be useful'.

**Cross-Activity Patterns:** Across all activities, Usefulness, Relevance, Flexibility, and Authenticity clustered at a median of 6 (IQR ≤ 1.25), indicating strong perceived instrumental value particularly in the prototype evaluation phase. Ratings for non-repetitiveness were more variable with the lowest median (4 in user research, rising to 6 in prototype evaluation) and a wide spread (IQR = 2), indicating that early interactions with the IVP were often experienced as repetitive, while later-stage tasks resulted in more varied output. This increase suggests the IVP is particularly effective when tasks involve clear boundaries and tangible design artifacts, allowing it to anchor responses concretely and avoid repetitive outputs. Conversely, open-ended user research appeared more challenging for the IVP, likely due to the difficulty of providing diverse yet realistic answers without concrete contextual anchors. This finding challenges assumptions about the seamless integration of AI personas throughout all design phases. Instead, it suggests that IVPs can be more effective in structured, artifact-based design stages, such as evaluating wireframes, rather than open-ended exploratory phases where nuanced perspectives are critical. Believability, the extent to which Alice could "pass" as a real person, showed wide dispersion (IQR = 2–3), revealing divergent comfort levels in treating the IVP as a credible user proxy. This variability is not merely aesthetic; it signals fundamental differences in how designers interpreted and engaged with the persona's outputs, a theme explored in more depth below.

**Experience-level Comparison**: When comparing subgroups based on UX experience: early-career junior (< 5 years; n = 4) and mid-career senior (≥ 5 years; n = 4), several differences emerged. Both juniors and seniors strongly affirmed the IVP's Usefulness (both Md=6), yet juniors showed tighter consensus (IQR=0.5 vs. seniors' IQR=1.25). Similarly, juniors rated Relevance slightly higher (Md=6, IQR=0.5) than seniors (Md=5.5, IQR=2), suggesting more experienced designers perceived broader contextual variability in IVP's suitability across design scenarios. The most notable subgroup difference emerged in perceptions of the IVP's authenticity and realism. Junior designers rated Real-Person Believability higher (Md = 6, IQR = 1.25) than senior designers (Md = 4, IQR = 4), indicating a 2-point gap with markedly tighter consensus among juniors. A similar pattern appeared for Authenticity, where juniors again showed more consistent approval (Md = 6, IQR = 0.25) compared to seniors (Md = 5, IQR = 1.5). This divergence suggests that professional experience acts as a moderating lens in evaluating AI personas. Junior designers who might benefit most from guidance are more susceptible to accepting AI responses uncritically, potentially missing the nuanced, messy reality of actual users. Furthermore, it suggests that senior designers' scepticism, while protecting against over-reliance, might cause them to underutilize IVPs. Both subgroups report mid-range Diversity (Md = 5–5.5) and similar spreads on Non-Repetitiveness (IQR = 2), indicating that repetitive outputs are a shared concern. It's important to note that with a small sample size (n=4 for each group), these findings are indicative rather than conclusive.

These quantitative findings provide an initial overview and backdrop for participants' perceptions. For instance, the consistent high ratings for 'Usefulness' and 'Flexibility' across tasks (Figure 1a) suggest a general positive reception. However, the variability in ratings for 'Non-repetitiveness' and 'Real Person' believability, especially the divergence between junior and senior designers' perception of 'Real Person' (Figure 1b), hints at specific challenges. While no inferential claims are made, these observed trends provide directional insights that complements and helps interpret the qualitative themes explored in the next section.

### 4.2 Participants' experience with Interactive Virtual Persona

Table 3 summarises the major themes that emerged from the analysis of post-task interview transcripts. Across five themes, three insights stand out: (i) the IVP is an effective early-phase collaborator that designers role-shift via prompts; (ii) voice adds presence yet reciprocity is limited, producing one-way exchanges; and (iii) over-optimism is persistent, so designers must actively solicit critique to avoid shallow trade-off exploration. We will discuss how each theme in relation to how it addresses **RQ1** (How LLM-powered IVPs support or hinder UX designers' practices) and **RQ2** (IVPs' perceived benefits and limitations) in the next subsections. The themes are presented in an order that progresses from those perceived as most supportive and beneficial to those seen as more hindering and limiting. Furthermore, we will integrate relevant survey data from Section 4.1 to illustrate points of convergence or nuance between the qualitative and quantitative findings.

**Table 3.** Major themes and how the themes address the RQ1: How IVPs support or hinder UX practice; and RQ2: IVPs' perceived benefit and limitations

| Themes | How the themes address RQ1 and RQ2 |
|---|---|
| Collaborative design partner | IVPs simulate stakeholder feedback, co-ideate design solutions, and evaluate wireframe prototypes (RQ2), supporting early-stage exploration and critique (RQ1). |
| Efficiency and practicality | By removing the need for participant recruitment, preserving conversational history across sessions, and enabling instant scenario exploration (RQ2), IVPs streamline logistics and facilitate efficient iterative design (RQ1). |
| One-way human-like interaction | While the human-like tone is perceived positively (RQ2), IVPs' limited reciprocity can hinder natural dialogue and collaborative inquiry (RQ1) |
| Biased, incomplete or inaccurate responses | IVP responses may reflect training-data biases and include hallucinations (RQ2), requiring designers to invest additional effort in validation, correction, and fact-checking (RQ1). |
| Over-optimism | IVP showed a consistent tendency to agree uncritically with design proposals (RQ2), suppressing critical feedback, inhibiting realistic constraint identification, and potentially misleading designers towards unviable design directions (RQ1). |

#### 4.2.1 *Collaborative design partner*

Participants highlighted the significant role of IVPs as a collaborative design partner throughout various stages of the design process, by assuming multiple roles. In many cases, IVPs acted as **a stakeholder**, offering a broad overview of user perspectives. P6S noted the utility of IVPs during the initial research phase and the ideation phase, particularly when designers lack familiarity with the user, as IVPs can "*give a broad kind of overview of who the user is*." As **a co-creator**, the IVP contributed fresh ideas and helped overcome creative blocks. P2J explained that the IVP helped generate innovative ideas, stating, "*generate some ideas on AI ... just see what that comes up with. You're like, Oh, I like that idea. Go along with it.*" Moreover, participants observed the IVP functioning as **a tester**, evaluating design concepts and providing feedback on wireframes. P1J remarked on the effectiveness of wireframe analysis, saying, "*I thought the wireframe analysis was really good. And her feedback on that was great*," highlighting how the IVP provided actionable insights at critical design junctures. Beyond ideation, participants found IVPs instrumental in accelerating the design validation process. P5S explained, "*We don't need to spend a lot of time trying to find a* *wide group of customers ... We can just go and chat straight away to someone like Alice and get the comprehension and the friction points understood quite quickly*."

Furthermore, the designers' specific prompts in their interaction with the IVP also showed how they assigned different types of roles to IVP through various speech acts across the three design activities (see Table 4).

**Table 4.** Varying IVP roles observed across three phases of design process

| User Research | IVP Roles |
| --- | --- |
| P6S: "Yeah. As a restaurant owner, what are some of the biggest challenges …?"<br>P3J: "Can you tell me a bit about your competitors? [...] what kind of people would you be competing around with?"<br>P1J: "I guess I just wanted to know a little bit more about your business. Can you tell me a bit more about it?" | Stakeholder / Business owner |
| **Ideation** | |
| P4J: "What are the features you think you require to tackle the above problems quite efficiently?"<br>P5S: "I feel like that's asking a lot of one tool… if you could prioritize… what would that be?"<br>P7S: "And I was also thinking for a mobile application… they can mix and match ingredients… What do you think about it?" | Co-creator / Collaborator |
| **Prototype Evaluation** | |
| P6S: "Is there anything that you think… is missing from any of these pages in terms of our overall goal?"<br>P8S: "Hi Alice, so we have an app… do you think if this is something you would like…?"<br>P2J: "Okay, what do you think of these wireframe UI images…?" | Tester / Evaluator |

Designers' prompts were performative. They did not merely elicit data; they created a context in which the IVP was expected to respond in a particular way through naturally occurring conversations. Taken together, these insights position IVPs as a multifaceted design partner that supports deeper user understanding, solution ideation, and preliminary evaluation. In essence, these varying roles undertaken by IVP illustrate the idea of "one persona, multiple roles" reflecting a dynamic approach where a single IVP adapts to different contexts, much like real-world stakeholders who acts as research participants, creative contributors, and evaluators at various stages of design process.

The perception of collaborative utility was supported by the survey findings (Section 4.1). For instance, 'Usefulness' (Md=6 across all activities) and 'Flexibility' (Md ranging 5 to 6 across all activities) were consistently rated highly (Figure 1a), indicating participants found the IVP instrumentally valuable as a partner. The high rating on 'Relevance' (median scores hovering around 6, particularly strong in prototype evaluation) further suggests the IVP's contributions were seen as pertinent to design tasks. While both junior and senior designers affirmed Usefulness (both Md=6), the tighter consensus among juniors (IQR=0.5) compared to seniors (IQR=1.25) (Figure 1b) might suggest that more experienced designers, while finding it useful, perhaps saw more contextual variability in its application as a 'partner' or held it to a higher standard of partnership.

Ultimately, the IVP's capacity to embody multiple roles can support UX designers' practices (RQ1) by offering on-demand, versatile assistance across different design stages, from initial user understanding to concept evaluation. Consequently, specific benefits (RQ2) include the potential for designers to rapidly gather diverse 'stakeholder-like' perspectives and creative input without the logistical overhead of engaging multiple human actors early on, thereby accelerating initial design explorations.

### 4.2.2 *Efficiency and practicality*

IVPs were also valued for their efficiency and practical utility, offering clear support for designers' workflows (RQ1) and delivering concrete benefits (RQ2). Participants consistently emphasized IVPs' ability to streamline the design process, remove logistical constraints, and enhance workflow continuity. Unlike traditional user research, which requires meticulous planning, scheduling, and participant recruitment, IVPs enable designers to simulate user interviews on-demand, making early-stage exploration faster and more flexible. In this sense, the IVP serves as a lightweight user research tool, enabling designers to explore user needs, motivations, and pain points interactively without the logistical constraints of involving real users in the early phases of design. P6S described this flexibility, explaining that IVPs "*remove the constraint of like having to be so organised*" compared to scheduling conversations with real users. Beyond accessibility, IVPs also enhance efficiency by preserving conversation history, allowing for a more fluid and iterative design process. P4J noted, "*I didn't feel that it is starting over from scratch. It was able to recall the previous prompts that we gave,*" highlighting how this persistent memory facilitates deeper exploration without requiring users to repeatedly provide context. This continuity saves designers valuable time, a benefit P4J further reinforced by stating, "*it will help us save so much time*."

Additionally, participants recognized IVPs' potential to extend beyond user representation to serve as adaptable design proxies for other stakeholders. P3J envisioned a scenario where an IVP could be customized to reflect a client's values and preferences, explaining, "*if I were able to train a particular ChatGPT model in a certain kind of way that … inhibits the basic core values … or the ideas of my client as well, then, it would be much more approachable, or you can say much more easier to me, easier for me to create all the designs, and like, have some interactive sessions, … have some brainstorming sessions … so that I can build everything that is like preferred by that particular client as well."* This suggests that IVPs could facilitate not just user-centric design but also internal alignment, helping teams prototype and refine ideas based on various stakeholder perspectives.

The survey data aligns with these qualitative insights on efficiency and resource-saving benefits. As noted earlier, the consistently high 'Usefulness' ratings indicate the practical value and efficiency benefits perceived by designers. Furthermore, the IVP's 'Flexibility' (Md ranging 5 to 6) and ability to provide 'Relevant' feedback (Md ranging from 5.5 to 6) across diverse tasks suggest its practical utility in varied design contexts, contributing to an overall more efficient and less resource-intensive design process.

Overall, IVPs can support UX designers' practices (RQ1) by reducing time and logistical burdens associated with traditional user engagement methods and enabling more continuous, iterative design explorations. The specific benefits (RQ2) derived from this include accelerated design cycles, on-demand access to user-like feedback, and the potential for broader contextual exploration without repeated setup.

### 4.2.3 *One-way human-like interaction*

Several participants (P2J, P4J, P5S, P7S) appreciated the conversational, natural tone of the AI persona, noting how it created a more engaging and human-like interaction. P2J described it as having a "*friendly voice*" that was "*trying to be a lot more intuitive in how we think*" while still maintaining the distinction between AI and human interaction. The persona's conversational tone contributed to an experience that felt more "*like talking to a second brain*" rather than a standard AI tool (P2J). Beyond tone, participants valued how the persona adapted to conversation flow, making the exchange feel more dynamic and engaging. In usability testing sessions, designers wanted to ask questions in a way that felt informal and comfortable, rather than scripted. One designer explained, "*I wanted to ask my questions in a way that felt conversational, because how we like to run our usability sessions is quite informally and making the participant feel comfortable. And, you know, our questions will change depending on the answers that the participant gives. So, I didn't want to feel like I was reading my questions off a sheet. I wanted it to feel like I was trying to have a conversation with her. And she did quite a good job at making it feel like a conversation*" (P5S). This remark reflects a desire for IVPs to not only respond accurately but also to adapt its questioning in a way that mirrors real-life human interactions. Additionally, subtle linguistic cues such as hesitations and filler words helped create a more human-like interaction. One designer observed how small verbal elements such as "um," "*I think*," or "*I pretend*" made the AI "*feel as human as possible*" (P7S). Another participant noted that the persona's speech was not only conversational but also incorporated "*jargon… relevant to her industry*", making responses feel "*natural and real*" while aligning with the domain of discussion (P5S).

Several participants (P3J, P4J, P5S, P8S) shared their experiences with the voice-based modality of the IVP, noting how it shaped their engagement across design activities. They highlighted how hearing Alice speak "very authentic… like you talk to the person, like a second brain" (P4J) and how the voice "didn't sound robotic, sounded friendly" (P4J) lowered the barrier to engagement and made impromptu check-ins feel more natural than typing. One participant noted that "as I started to get used to Alice's voice … I felt a lot more comfortable compared to the beginning" (P5J), suggesting the quick acclimatisation that voice afforded. At the same time, participants also identified limitations related to conversational flow and unidirectional interaction. Responses were occasionally too lengthy and difficult to interrupt with one participant noting "very time consuming… I would have preferred a bit smaller or shorter answers" (P3J), and the lack of reciprocal cues led one senior designer to feel it was "a one-way conversation… I'd like an exchange of ideas" (P8S). Furthermore, P1J noted, "in ideation… you don't want the other person to talk for a long time. You want them to just go like idea, idea, idea," and P2J remarked, "Alice can't really read my mannerisms. Can't really tell … if I'm like implying something, or … if I have a certain tone of voice… I have to wait for her to finish." Without IVP picking up on real-time non-verbal cues or interjecting with its own curiosity, participants felt that much of the dialogue was a "one-way street." Furthermore, the IVP's tendency to generate lengthy responses was noted to reduce user engagement as mentioned by P3J – "*it gave me a list of 20 different something like that. It was engaging up till that point*". These insights suggest that while the voice modality offers potential for more engaging and embodied interaction, its utility could be improved by enabling configurable response lengths, user interruptions, recognition of non-verbal communication cues, and greater conversational reciprocity.

The survey results reflect this nuanced experience. While 'Authenticity' generally received positive median scores (Md=5 for seniors, Md=6 for juniors; Figure 1b), the extent to which Alice could "pass" as a 'Real Person' showed greater variability (IQR = 2-3 across activities, Figure 1a) and an indicative subgroup trend, with juniors reporting higher believability (Md = 6) than seniors (Md = 4; Figure 1b). This suggests that while the tone was human-like (contributing to 'Authenticity'), the lack of true reciprocity may have impacted its 'Real Person' believability, particularly for senior designers. Seniors appear to have associated "real" with dialogic reciprocity and constraint-aware critique, describing the exchange as "one-way" and "self-affirmatory," and noting that Alice never treated ideas as infeasible, a pattern we explore further in the Over-optimism theme (4.2.5). Juniors, by contrast, tended to value human-like tone and immediacy while still wanting briefer turns or occasional argumentation to feel more believable.

Overall, the human-like conversational tone was positively perceived (RQ2); however, the predominantly one-way nature of the interaction ultimately hindered the development of a truly reciprocal dialogue and collaborative inquiry (RQ1). While an initial ease of interaction and a more engaging experience were perceived useful, the reduced sense of genuine conversational partnership, an inability for the IVP to proactively guide or clarify, and potential for user disengagement with overly verbose, non-interactive responses were major limitations (RQ2).

#### 4.2.4 *Bias, incomplete or inaccurate responses*

A significant concern revolved around the potential for bias and inaccuracies stemming from the underlying LLM. P6S explicitly raised concerns about "*bias of information or lack of completeness of information*." This is aligned with a fundamental challenge with LLMs: they are trained on vast datasets that may reflect existing societal biases (Ferrara, 2023). In the context of IVPs, this can manifest as biased representations of user groups or skewed perspectives. The fact that the IVP was based on a single hypothetical user persona was also seen as a limitation, as P6S noted the concern of "*not very diverse sample size*". This suggests that relying on a single IVP may not adequately capture the diversity of user needs and perspectives within a target audience. Further, P6S questioned the depth of information provided, describing it as a "*distilled set of information rather than … something truly insightful*," raising concerns about the IVP's ability to provide truly nuanced and insightful feedback compared to real user interaction. P2J directly questioned the representativeness of the IVP, asking "*whether the IVP truly represents the actual target audience*." This is a critical validity concern, as the effectiveness of personas, virtual or static, hinges on their accurate representation of target users.

Participants also noted inconsistencies in the quality and depth of IVP responses. Several instances were reported where the IVP lacked depth, misinterpreted wireframe features, or provided irrelevant suggestions. Specifically, participants observed that "*the model was not able to see some parts of the wireframes which were already there, and, in some cases, it was explaining the features which were not there*." This points to limitations in the IVP’s ability to process and understand visual information, particularly in the context of wireframe analysis. Furthermore, concerns about repetitiveness were voiced, with P8S stating, "*But sometimes, she was just repeating a lot of information*." This repetition, also reflected in the survey data's lower scores on "non-repetitiveness," reduces the perceived value and engagement with the IVP. A more fundamental concern was the potential for the IVP to prioritize known solutions over innovative ones. As P6S articulated, "*having a language model prioritise known solutions, rather than uniqueness of ideas and originality of concepts*" could stifle creative exploration and limit the generation of truly novel design solutions.

Notably, many participants attributed these shortcomings not only to the IVP itself but also to their own lack of skill in crafting effective prompts. P8S noted, "*I feel that what could be improved in this, maybe I could try and frame my questions in a better way… ask more specific questions and see how she answers*." Similarly, P1J admitted, "*I probably needed to ask better questions,*" while P4J reflected, "*I think I can ask … more crisp questions*." P5S summarized this learning curve by stating, "*the more you use it, the better you would get at knowing how to ask questions*." This suggests that while the IVP had inherent limitations, its effectiveness was also influenced by users' ability to refine and iterate on their prompts. Many realized that obtaining useful responses required significant effort in adjusting their queries, further emphasizing the interactive and skill-dependent nature of working with an IVP.

The survey data largely corroborate the concerns voiced in the interviews. Across all three activities, "Non-repetitive" and "Diversity" attract the lowest medians, signalling that the participants frequently perceived the IVP as recycling content and failing to surface a broad range of perspectives. In contrast, the high medians for "Relevant" and "Useful" (around 6 across all tasks) substantiate comments that, despite bias and depth issues, most designers still judged the IVP's answers to be practically valuable even when they doubted its depth or representativeness.

Overall, concerns over inherent biases, limited user representation, superficial and sometimes repetitive insights, and misinterpretations of visual elements (RQ2) suggest that the IVP was unable to effectively simulate the diverse and nuanced needs and perspectives of the actual target user group in their full richness, requiring designers to invest additional effort in validation, correction, and fact-checking (RQ1). Moreover, participants recognized that the IVP's effectiveness was not solely constrained by its own limitations but also by their ability to refine and adapt their prompts. Many found that obtaining meaningful and contextually relevant responses required ongoing adjustments to their queries, with some acknowledging that their initial lack of prompting skills contributed to poor response quality.

#### 4.2.5 *Over-optimism*

One of the most consistent observations across participants (P3J, P6S, P7S, P8S) was that IVP did not push back against ideas, unlike real-world users, who often highlight limitations, doubts, and frustrations. Instead, IVP's responses were overwhelmingly optimistic and accommodating, rarely acknowledging challenges such as technical feasibility, cost, or competing priorities. "*No that's not possible,*" one participant noted, reflecting on how IVP seemed to accept every proposal without hesitation, rather than engaging critically with its realistic implications (P7S). Another participant expressed similar concerns: "*It felt like every idea I threw at her was feasible… but in reality, things don't work that way*" (P8S). The following extracts illustrate the IVP's over-optimistic and non-critical approach:

> **P8S**: Hi Alice, do you think having a two-hour session about organic farming in your own restaurant will help you gain more customers?
>
> **Alice (IVP)**: Yeah, that could be really effective… It's a great way to engage directly with customers.

While not wrong, Alice gives a wholehearted "Yes" without mentioning practical hurdles—staffing needs for such events, scheduling constraints, costs, or how often such sessions can realistically occur. She does not raise any downside or logistical concern.

> **P2J**: What if there was some sort of gamification… or a loyalty program that rewards returning the packaging?
>
> **Alice (IVP):** Ahh, I love that idea! Gamification could really amplify engagement… It's a fun way to strengthen the community around our sustainability values.

Alice praises the concept with enthusiastic "*love that idea!"* but never questions real-world complexities of launching a full gamification system (e.g., cost, user adoption challenges, tech resources needed). The response stays squarely positive. This "over-optimism" and absence of simulated constraints had several practical impacts on the design process as observed by participants:

- **Reduced Identification of Design Flaws and Unchallenged Assumptions**. Alice's agreeable nature meant designers received less inherent pushback. As P6S stated, "*often when I'm doing usability testing… you get quite a lot of negative feedback which is often really quite helpful*." Alice's inability to spontaneously offer this critical, negative feedback, or to simulate the scepticism a real user might have towards a new feature or complex interaction, could lead to designers proceeding with less robust or unrefined ideas without early critical scrutiny.

- **Potential for Over-Trust or Misguided Confidence**. The consistent positive reinforcement, especially noted by P8S's reflection that "*according to her, everything was feasible*," risked designers, particularly those less experienced, placing undue confidence in initial concepts. P1J explicitly voiced a concern about "trusting it [Alice] too much." This could lead to an overestimation of an idea's viability or user appeal, potentially resulting in misallocated design efforts towards features that a real user or stakeholder would have quickly deprioritized due to practical constraints.
- **Superficial Exploration of Constraints and Trade-offs**. Because Alice did not embody real-world limitations such as budget and time, discussions around complex issues like business strategy or feature prioritization often lacked depth. When P7S probed about funding for expansion, Alice offered optimistic, high-level suggestions like exploring external funding options, rather than engaging with the specific, granular financial challenges a real business owner would articulate. This meant designers were not forced to make the difficult trade-offs and compromises that characterize authentic design problems.
- **Inability to Simulate Genuine Emotional or Contextual Friction**. Real users express frustration, confusion, or impatience, which are vital cues for design improvement. P6S highlighted that real users, "*when they're really annoyed... tell you that like 'I just hate it'.*" Alice could not replicate this emotional friction. For example, if a proposed app flow was cumbersome, Alice would process it logically based on the provided information but would not convey the impatience a busy user might feel, thereby missing an opportunity for the designer to identify and address a key usability issue.

P3J also pointed out that the sheer volume of features suggested by the IVP, "*it gave me a list of 20 different something like that*," could lead to feature overload and potentially negatively impact user experience if all suggestions were implemented without careful consideration. This suggests that while IVPs can be generative in ideation, they may lack the critical judgment necessary to filter and prioritize design ideas based on feasibility and user needs, potentially leading designers down impractical or less effective paths. This sycophantic tendency, where the AI model tends to agree and generate positive responses, has been observed in LLMs and requires careful consideration in their application as personas (Sharma et al., 2023).

The lack of critical feedback is indirectly reflected in the survey's 'Authenticity' and 'Real Person' believability scores. While participants found the IVP somewhat authentic in its persona (Md=5-6 for Authenticity, Figure 1a/b), the consistent over-optimism would likely detract from its believability as a realistic user proxy, especially for experienced designers (Seniors Md=4 for 'Real Person', Figure 1b). Real users are rarely so uniformly positive. The survey did not directly measure "criticality," but the qualitative evidence strongly suggests that this over-optimism limits the IVP's utility in simulating the pragmatic tensions and critical feedback essential for refining design ideas, thereby hindering a more grounded design exploration (RQ1). In real-world user sessions, participants may push back, express frustration, get confused, and even contradict themselves. Without this real-world friction (RQ2), IVP risked producing polished but unchallenged design directions.

Overall, across all design activities, participants' perceptions of the IVP varied according to experience level and task type. Senior designers generally regarded the IVP as a useful but supplementary tool, valuing its ability to provide broad overviews and quick feedback, particularly in tasks such as prototype evaluation. However, they expressed reservations about the lack of critical engagement, noting that the IVP's responses were often overly agreeable and failed to reflect realistic constraints or user skepticism. Junior designers, by contrast, exhibited greater initial enthusiasm, especially in ideation and early-stage research activities. They valued the IVP's immediacy and generative capacity, despite recognizing its limitations. Consensus was strongest around the utility of the IVP in prototype evaluation, which was widely viewed as its most functionally supportive role. Nonetheless, concerns about over-optimism were raised across both groups, with four participants observing that the IVP frequently affirmed design ideas without prompting deeper reflection or surfacing potential drawbacks. Ultimately, the IVP was generally seen as useful, especially for ideation and rapid prototype feedback, but its pronounced over-optimism and the need for real-user validation, particularly in user research, and remained key considerations for many participants.

**Answer to RQ1: How do LLM-powered IVPs support or hinder UX designers' practices?** IVPs supported practice by facilitating on-demand, context-preserving collaboration across user research, ideation, and prototype evaluation, while also reducing recruitment and scheduling overhead. At the same time, they hindered practice through one-directional and verbose exchanges, occasional repetition and misinterpretation of visual artefacts, and limited critical pushback, necessitating human validation.

**Answer to RQ2: What are LLM-powered IVP's perceived benefits and limitations?** Designers valued efficiency, on-demand access, and the naturalness of the voice interaction, with IVP's utility strongest in prototype evaluation. However, limitations included concerns regarding bias and representativeness, shallow or repetitive responses, variability in perceived "real-person" believability (typically higher among junior than senior designers), and a tendency toward over-optimism. Overall, IVPs were valued as complementary tools that accelerate early design cycles and prototype feedback, but not as substitutes for engagement with real users.

## 5. DISCUSSION

### 5.1 Amplifying the potential and mitigating the perils

**Value across design phases and various roles**. The results indicate that IVPs are perceived as a valuable collaborative design partner undertaking different roles including business owner, co-creator, and evaluator. Designers found IVP useful for initial user research, brainstorming, and gaining rapid feedback on design concepts. This aligns with previous research highlighting the potential of GenAI to accelerate various aspects of interactive system development (Schmidt, 2023) and support tasks like creating consumer profiles and assuming different roles in the design process (Hong et al., 2023; Kocaballi, 2023). This capacity to offer on-demand user engagement creates an opportunity to enrich the early design phases with human-centred thinking, particularly when real users are not immediately available.

**Strategic integration across design phases.** As shown in Section 4.2.2, the efficiency and accessibility of IVPs were particularly appreciated, allowing designers to quickly explore user needs and pain points without the time and resource constraints of traditional user research methods. However, realizing this potential requires a strategic and nuanced approach. It is not merely a matter of replacing traditional personas with their interactive counterparts. Instead, it involves strategically integrating IVPs into existing workflows in a way that complements, rather than replaces, established methods such as user interviews and usability testing (Zhou et al., 2024). Designers can integrate IVPs at junctures where rapid prototyping, initial feedback, or broad conceptual brainstorming can benefit from *user-like* perspectives. Rather than relying on IVPs universally, selective incorporation in high-level brainstorming, early user research, or low-fidelity prototyping tasks can improve efficiency and bolster creative momentum while preventing over-reliance on potentially ungrounded and superficial insights.

**Effective prompt engineering**. The quality of IVP feedback was intrinsically linked to the specificity and clarity of the prompts as noted by participants (P1J, P4J, P5S, P8S). Targeted questions can elicit more nuanced, contextually relevant responses and help mitigate irrelevant outputs or issues like LLM sycophancy (Sharma et al., 2023). Simple yet effective strategies include specifying the desired level of detail in the system prompt, introducing constraints (e.g., "focus your responses on cost-feasibility concerns"), and iteratively refining prompts as the design evolves and as informed by prior responses. However, an overemphasis on explicit prompt engineering can disrupt the natural flow of conversational interaction. Rather than requiring designers to master advanced techniques like few-shot learning (Song et al., 2023), these principles can be embedded within the IVP's system design or persona specifications. This approach can allow designers to engage conversationally while still receiving focused and useful feedback. Our study did not focus on the creation of effective LLM-based personas. However, recent work offers a structured data-driven method to creating LLM-based personas (Jung et al., 2025). The study found LLM-based personas outperformed those created without LLM assistance across several criteria including clarity, completeness, consistency, credibility, and fluency. However, the study's focus was on the creation of static personas. Therefore, the method's effectiveness in creating LLM-based interactive personas requires further empirical research.

**Bias awareness and human-in-the-loop engagement**. Despite IVPs' appropriate responses in many cases (Sections 4.2.1 and 4.2.2), participant reflections supported a human-in-the-loop paradigm (Section 4.2.4). As Friess (2012) and Matthews et al. (2012) caution, personas, digital or otherwise, risk fostering superficial understanding unless designers critically engage with their content and outputs. In our study, designers actively evaluated and contextualized IVP suggestions, preventing over-reliance on AI. This aligns with broader calls for collaborative AI approaches in HCD (D. Wang et al., 2020; Weisz et al., 2024), where humans interpret and validate AI outputs rather than passively accepting them. Designers must be trained to recognize the limitations of LLMs, including their potential for bias and hallucination, and to approach IVP-generated insights with an appropriate level of scepticism.

Although there are several mitigation strategies against LLM biases such human-in-the-loop engagement, prompting strategies, bias audits, recent evidence shows that LLM biases can be deep-rooted, hard-to-discern, and resistant to prompt engineering (Gupta et al., 2024).

**Managing over-optimism**. As reported by four participants (Section 4.2.5), our study found that IVPs tended to present all design suggestions as feasible, a pattern that risks overlooking practicality and real-world constraints. This is in fact one of the shortcomings of LLMs using "Deep Reinforcement Learning from Human Preferences" (RLHF) (Christiano et al., 2017), a behaviour known as sycophancy in LLMs (Ranaldi & Pucci, 2024; Sharma et al., 2023). RLHF improves models by aligning them with human feedback, but such alignment can be prone to biases based on the type of feedback provided. For instance, if evaluators show a bias toward rewarding responses that encourage user queries (even when unrealistic), the model could be conditioned to respond as if all queries are feasible (Christiano et al., 2017). The debate over LLM sycophancy gained widespread public attention when OpenAI was forced to roll back its latest version of ChatGPT-4o due to user reports of excessive sycophantic behaviour (*Sycophancy in GPT-4o*, 2025). LLMs' over-optimism or sycophancy may mislead inexperienced users into pursuing impractical ideas. We advocate positioning IVPs as idea generators rather than decision-makers. Designers can impose feasibility checks, prompting the IVP for explicit cost-benefit analyses or requesting risk-based critiques. By systematically soliciting "devil's advocate" viewpoints, teams can counterbalance the persona's innate positivity. While mitigating sycophancy in LLMs is still an open research area, there are emerging evidence on some partially effective methods such as combining prompt engineering with in-context exemplars (RRV et al., 2024) and synthetic data augmentation (Wei et al., 2024).

**Ethical implications of AI-simulated users.** Participants raised clear ethical risks including over-trust in plausible outputs when feasibility limits were hidden (Section 4.2.5) and lack of representational diversity in a single persona (Section 4.2.4). The ethical dimension of simulating users, especially as technology edges closer to realistic persona embodiments, cannot be overlooked (Aher et al., 2023; Schmidt et al., 2024). While simulated personas can jumpstart early ideas, they must not supplant genuine user voices, particularly for sensitive or marginalized populations. It is crucial to avoid creating a false sense of empathy or anthropomorphizing these tools in a way that could lead to designers neglecting real user research (De Freitas et al., 2023; Placani, 2024). Transparent disclosures about AI-driven personas (Salminen, Jung, et al., 2020), ethical guidelines for data usage (Ryan & Stahl, 2020), and careful moderation of IVP outputs are vital for maintaining the integrity of human-centred values. The ethical implications of using IVPs must be continuously debated and addressed through guidelines, best practices, and ongoing dialogue within the HCI research and design communities.

### 5.2 Towards a balanced approach: A Preliminary Framework for Responsible Use

Together, these opportunities and challenges indicate that IVPs can be embraced as augmentative tools within a broader design ecosystem. Building on the concept of "human-AI co-creation" (Wu et al., 2021), our findings indicate that IVPs are most effective when designers maintain full control of the process, carefully curate prompts, interpret feedback, and remain accountable for ethical decisions. When designers use IVPs as brainstorming partners or as quick user-proxy checks while maintaining their role as arbiters of design judgment, IVPs can augment creativity and efficiency without overshadowing human intuition and empathy. Human judgment, empathy, and contextual awareness remain critical for identifying when the IVP's suggestions drift from user realities. While IVPs can support creativity and accelerate early-stage design, they must not overshadow stakeholder involvement or user interviews.

Drawing on our key findings, we propose a **preliminary framework for responsible IVP use** comprised of five principles: ethical transparency, bias vigilance, critical evaluation, complementary research, and iterative refinement (Table 5).

**Table 5.** A preliminary framework for responsible use of IVPs

| Principles | Study Findings | Key insights and Prior Research |
|---|---|---|
| **Ethical Transparency** | The IVP, Alice, was often anthropomorphized, referred to as 'she,' and described as having personable, human-like traits, with high human-likeness scores (50%-75%). If unqualified, the simulated authenticity of responses risks being misinterpreted as genuine user input. | Disclosing the AI-driven nature and inherent limitations of IVPs is critical. Transparency ensures that stakeholders understand that the persona is a simulated construct, thereby preventing misinterpretation and safeguarding design decisions from being based on assumed human authenticity. Prior work highlights transparency as key to human trust (Diakopoulos, 2016; Mills, 2024), offers guidelines for disclosing system limitations ("IEEE Standard for Transparency of Autonomous Systems," 2022) and proposes |

| Principles | Study Findings | Key insights and Prior Research |
|---|---|---|
| | | "model cards" that detail model characteristics and limitations (Mitchell et al., 2019). |
| **Bias Vigilance** | There were concerns that (i) a single IVP might incorporate narrow, culturally skewed perspectives, and (ii) the limited representativeness could reinforce existing biases. | Continuous auditing and proactive mitigation measures are necessary to address the possibility of various bias. By regularly reviewing and updating the IVP's data sources and configuration, designers can ensure that the simulated persona reflects a broader range of user perspectives and avoids stereotyping. Prior work shows LLMs (Guo et al., 2024) and LLM-generated personas (Salminen et al., 2024) can have gender, age, cultural and occupational biases, and LLM simulations can oversimplify human behaviour into caricatures (Cheng, Piccardi, et al., 2023). |
| **Critical Evaluation** | The IVP's outputs were found over-optimistic, superficial, repetitive, and lacked the depth of real user feedback. This limitation necessitated additional human scrutiny to interpret the results. | Rigorous human oversight is essential to critically assess AI outputs. Designers should use established heuristics, domain expertise, and real user data to evaluate and contextualize the IVP's feedback, ensuring that design decisions are informed by nuanced and validated insights rather than by unfiltered AI responses. Prior work emphasizes the need for critical evaluation of automated outputs to prevent over-reliance , the importance of domain expertise in scrutinizing AI suggestions (Lai et al., 2021) and AI persona generation (Salminen et al., 2024), and proposes a model for Appropriateness of Reliance (Schemmer et al., 2023). |
| **Complementary Research** | Participants stressed that while the IVP expedites early ideation, it should not replace in-depth direct user research. The consensus was that genuine user engagement remains indispensable for reliable design. | The IVP should be used as a complementary tool alongside traditional user research methods. This dual approach guarantees that while IVPs provide rapid, on-demand feedback, especially in early design phases, the authenticity and complexity of human needs are captured through direct interviews and field studies, leading to more robust design outcomes. Prior work shows LLMs struggle with simulating subtle, contextual, and empathetic aspects of human experience (Gerosa et al., 2023) and require direct user insights for effective use (Gu et al., 2025). |
| **Iterative Refinement** | Participants highlighted the need to continually adjust prompts and system instructions to obtain more | An iterative refinement process involves continuously adjusting prompts and recalibrating persona specifications to generate more relevant responses. As design requirements and user insights evolve, regular prompt engineering, persona and data source refinements help ensure that the IVP |

| Principles | Study Findings | Key insights and Prior Research |
|---|---|---|
| | nuanced, adaptive outputs from the IVP. | remains aligned with current objectives and delivers relevant feedback over time. Modular, data-driven architectures and workflows (McGinn & Kotamraju, 2008; Jung et al., 2025) for creating and managing personas can facilitate the efficient updating and fine-tuning of persona specifications and dialogue management components independently. Emerging methods are also proposed to help steer LLMs toward desired personas (Li, Mehrabi, et al., 2024). |

This framework outlines a set of interdependent principles aimed at ensuring that LLM-based IVPs are deployed responsibly within HCD processes. The approach foregrounds ethical transparency, bias vigilance, critical evaluation, complementary research, and iterative refinement as essential for aligning automated persona outputs with genuine human experiences.

**Principle 1: Ethical transparency**

Participants frequently ascribed human-like traits to the IVP, "Alice," using pronouns such as "she," and referencing her authentic-sounding voice. This tendency to anthropomorphize the IVP points to a risk that the simulated authenticity of LLM-based personas may be confused with genuine user input, particularly when these personas yield responses with high human-likeness scores. To mitigate this, designers must be transparent about the AI-driven nature of IVPs. Transparency means explicitly disclosing that the persona's outputs derive from computational processes, clarifying its inherent limitations, and providing accessible documentation of its data sources, model architecture, and capability boundaries (e.g., "model cards" (Mitchell et al., 2019)). Such disclosures are critical for preventing stakeholder misunderstanding and ensuring that design decisions do not rely on assumed human authenticity. This aligns with existing work emphasizing openness about system limitations as a cornerstone of user trust and safety (Diakopoulos, 2016; Mills, 2024), as well as guidelines that formalize disclosure practices in autonomous systems ("IEEE Standard for Transparency of Autonomous Systems," 2022).

**Principle 2: Bias Vigilance**

A recurring concern among study participants was the potential for IVPs to reinforce or introduce biases due to their underlying training data and modelling approaches. Specifically, two key issues were identified: (i) the risk of incorporating culturally skewed perspectives and (ii) the limited representativeness of the IVP, which could lead to an overly narrow, homogenized user viewpoint. Addressing these concerns necessitates a proactive approach to bias mitigation, including continuous auditing of IVP responses and systematic refinement of data sources. Prior research highlights that LLMs and their applications in persona generation are susceptible to biases related to gender, age, culture, and occupation (Guo et al., 2024; Salminen et al., 2024). Additionally, studies have demonstrated that LLM-generated personas may inadvertently reduce human behaviours to oversimplified caricatures (Cheng, Piccardi, et al., 2023). To counteract these tendencies, a group of diverse personas, iterative updates and diversified training inputs are advocated for IVP workflows, ensuring that personas reflect a broader spectrum of real-world users.

**Principle 3: Critical Evaluation**

Study participants found that the IVP's outputs were overly optimistic, occasionally repetitive, and lacked the nuanced detail that real users provide. These limitations highlight the necessity of rigorous human oversight in evaluating AI-generated insights. Designers must apply domain expertise, established heuristics, and empirical user data to critically assess IVP outputs, thereby ensuring that AI-driven feedback is contextualized appropriately within the design process. Over-reliance on automated outputs without adequate scrutiny can lead to superficial or misleading conclusions, as documented in prior studies on automation bias (Parasuraman & Riley, 1997). The importance of human expertise in refining AI-generated suggestions has been further emphasized in AI-assisted decision-making research (Lai et al., 2021) and AI persona development (Salminen et al., 2024). The Appropriateness of Reliance framework (Schemmer et al., 2023) shows that effective human-AI collaboration depends on distinguishing between when users correct their initial errors by following accurate AI advice (relative AI reliance) and when they ignore flawed advice (relative self-reliance). In practice, designers must adopt a critical stance, validating AI-driven outputs against real user behaviours to ensure robust and meaningful design outcomes.

**Principle 4: Complementary Research**

While IVPs offered an efficient means of gathering rapid, on-demand feedback during early design stages, study participants consistently emphasized that IVP's feedback does not replace the depth and authenticity of real user research. Rather, designers viewed the IVP as a complementary instrument that can offer practical, low-cost insights in early stages of design, before undertaking in-depth interviews, ethnographic observations, or other rigorous methods. This hybrid approach capitalizes on the efficiency of AI-generated input while preserving the complexity, empathy, and cultural specificity gleaned directly from human participants. Research has demonstrated that LLMs struggle to accurately simulate nuanced social and emotional dimensions of human interaction (Gerosa et al., 2023), highlighting the necessity of direct user input for refining AI-driven insights (Gu et al., 2025). By integrating IVP-driven feedback with empirical user research, designers can leverage the efficiency of AI-generated feedback while maintaining the depth and authenticity of user-centred inquiry, leading to more holistic and reliable design decisions.

**Principle 5: Iterative Refinement**

Participants' experiences with and reflections on overly generic or superficial outputs highlighted the need for an iterative refinement approach. Key strategies for improving IVP effectiveness include fine-tuning prompts, recalibrating persona specifications, and curating data sources to better align LLM-generated responses with real-world user behaviours. This iterative process ensures that IVPs provide increasingly accurate and contextually relevant feedback over time. This refinement approach aligns with data-driven persona development frameworks (McGinn & Kotamraju, 2008; Jung et al., 2025) that can support modular reconfiguration and updates of persona specifications and data sources. While adjusting persona parameters and data inputs through a modular architecture can enhance IVP response quality, prompt engineering can play a particularly critical role in steering LLMs toward targeted persona goals (e.g., (Li, Mehrabi, et al., 2024)). By positioning IVPs as dynamic design tools rather than static representations, design teams can obtain more nuanced, contextually relevant insights that better reflect the complexity and richness of real-world user behaviours and expectations.

Our empirical findings lay an initial foundation for this preliminary framework, but they do not yet offer a comprehensive or finalized model. The identified challenges such as over-optimism, bias, and the need for critical evaluation are addressed through principles like Ethical Transparency, Bias Vigilance, and Critical Evaluation, while the practical advantages of IVPs, including rapid ideation and efficiency, highlight the relevance of Complementary Research and Iterative Refinement. However, given the limited scope and sample of our study, this framework should be regarded as a starting point, subject to further refinement, expansion, and validation. As future research explores IVPs across different domains, design contexts, and extended workflows, these principles will require continuous reassessment to ensure their effectiveness in guiding responsible and ethical IVP integration into HCD.

To further operationalize these principles, Table 6 presents some implementation strategies that designers can adopt to integrate IVPs responsibly into their workflows. It is important to note that, given the preliminary nature of our framework and the evolving landscape of LLM-powered design tools, these strategies are intended as a starting point and may require adaptation based on specific project contexts, IVP capabilities, and further research. Since this is a rapidly evolving research and application area, it is essential to follow emerging evidence on best practices.

**Table 6**. Implementation strategies for responsible use of IVPs

| Principle | Implementation Strategies |
|---|---|
| **Ethical Transparency** | • **Disclose AI involvement**: Clearly label all IVP-generated outputs as "AI-Simulated" and explain its non-human nature to all stakeholders (Diakopoulos, 2016).<br>• **Provide model cards:** Create one-page “model card” (Mitchell et al., 2019) with capabilities, known limitations, and intended use cases.<br>• **Explain responses:** Offer mechanisms for designers to understand why an IVP provided a certain response (e.g., key influencing factors from its persona profile or recent conversation). See (Salminen, Santos, Jung, et al., 2020) for implementing transparency in data-driven personas.<br>• **Improve capability**: Provide training on the IVP's specific limitations (e.g., no lived experience, potential for bias, hallucination, and sycophancy). |

| Principle | Implementation Strategies |
| --- | --- |
| **Bias Vigilance** | • **Create diverse persona sets**: Develop and utilize a portfolio of IVPs reflecting diverse user characteristics and backgrounds, as aligned with the original persona motivations (Cooper, 1999; Pruitt & Adlin, 2005).<br>• **Arrange bias audits**: Regularly test IVP responses against diverse scenarios and known bias types (e.g., gender, age, culture). Raji et al.'s (2020) auditing framework can be adapted for IVPs.<br>• **Curate training data**: If feasible, use or fine-tune models on datasets carefully curated to minimize societal biases. Solaiman & Dennison (2021) proposes creating "values-targeted datasets" to fine-tune LLMs. These datasets could be constructed to encourage desirable behaviors such as fairness and discourage undesirable ones such as bias. |
| **Critical Evaluation** | • **Mandate human review**: Establish protocols requiring human oversight to critically review, validate, and contextualize IVP outputs before design decisions are made. The protocol, for instance, may provide relaxed oversight requirements for low-stakes activities such as divergent ideation.<br>• **Perform cross-validation**: Triangulate IVP feedback with other data sources (e.g., analytics, prior user research, and expert reviews).<br>• **Support "appropriate reliance":** Provide training to discern when to trust IVP input versus their own expertise or other data. See (Schemmer et al., 2023) to learn about appropriate reliance in human-AI collaboration.<br>• **Teach prompt engineering**: Provide training to effectively use prompt engineering techniques to deal with the model limitations. |
| **Complementary Research** | • **Define usage scope and guidelines**: Clearly position IVPs as tools for early exploration, hypothesis generation, or low-fi prototype feedback, not as substitutes for direct user data.<br>• **Make validation essential**: Integrate IVP use in professional workflows involving validation (e.g., use IVP for initial ideation, then validate with real users).<br>• **Establish conflict resolution processes**: In cases of conflict, for instance, insights from direct user research can supersede IVP outputs. |
| **Iterative Refinement** | • **Provide feedback mechanisms**: Implement processes for designers to provide feedback on IVP performance (e.g., quality of critique, relevance, and bias).<br>• **Create prompting libraries**: Develop and share libraries of effective prompts designed to elicit specific types of feedback, including critical evaluation and friction. |

| Principle | Implementation Strategies |
|---|---|
| | • **Perform regular updates**: Periodically review and update IVP persona specifications, data sources, and system prompts based on evolving design needs and feedback. |

### 5.3 Comparison with previous work

Prior research has demonstrated the potential of LLMs in generating user-like content and supporting various design tasks (e.g., (Hong et al., 2023; Kocaballi, 2023)). Recent studies on LLM-powered interactive personas have explored: (i) single-phase applications, such as ability-based persona creation through text-based interfaces (Sun et al., 2024a); (ii) embodied interactive personas, exemplified by (Zhou et al., 2024) "Vivid Personas", which prioritize realism and technical feasibility; (iii) comparative evaluations of text-based interactive synthetic personas (Gu et al., 2025), and (iv) users' susceptibility to hallucinated answers, highlighting the need for a greater degree of transparency (Kaate et al., 2025). Furthermore, recent systems such as Proxona (Choi et al., 2025) and AIdeation (W. F. Wang et al., 2025) illustrate the evolving roles of LLM-powered personas: Proxona generates multi-dimensional, data-grounded audience personas to aid creator sensemaking and content refinement, while AIdeation facilitates iterative visual and textual ideation through a multimodal interface.

Within this evolving landscape, our study with its LLM-powered, multimodal, voice-enabled IVP offers specific points of differentiation when compared to existing approaches:

- **Compared to traditional static personas**: Our study empirically demonstrates how the interactive, conversational nature of the IVP can overcome inherent limitations of traditional static personas. While traditional personas can be one-way and impersonal, restricting ongoing dialogue and engagement (Section 2.1), designers in our study could directly interview, brainstorm with, and gather feedback from the IVP in real time via voice interface (Section 4.2.1). This dialogic engagement expedites information gathering and allows for spontaneous probing, providing a notable perceived advancement over the "passive reading" associated with static documents (Section 4.2.2). The IVP's ability to maintain conversation history further supported continuity of design activities and iterative exploration, contrasting with the often-underutilized nature of traditional personas (Section 4.2.2).

- **Compared to the other AI Assistants and General-purpose LLMs**: While GenAI can accelerate system development (Schmidt, 2023) and LLMs can create consumer profiles (Hong et al., 2023) or assume versatile roles (Kocaballi, 2023), our study highlights the unique value of a dedicated, simulated persona. "Alice" provided a consistent, character-driven interaction across an end-to-end design process (user research, ideation, prototype evaluation (Section 4.2.1). This contrasts with general-purpose LLMs which, while exhibiting a default interaction style, lack the explicitly defined and richly detailed persona characteristics of an IVP like "Alice". Our findings showed designers could performatively shape Alice's role through their prompts (Table 3, Section 4.2.1), eliciting stakeholder, co-creator, or tester perspectives from the same IVP. This demonstrates a more integrated, active, and sustained form of persona engagement than re-prompting a general AI assistant. The voice-based, multimodal interface of the IVP also supported more naturalistic interactions (Section 4.2.3)
- **Compared to other interactive and AI Personas**: Our work builds upon and differentiates itself from recent AI persona studies by (i) demonstrating the end-to-end application of a voice-first, multimodal IVP, highlighting its perceived naturalness and cross-activity utility with consistently high usefulness and flexibility ratings across multiple design activities; (ii) identifying the persistent "over-optimism" of an LLM-driven persona as a specific, recurring challenge for designers seeking critical feedback, contributing to the growing body of research showing the implications of general LLM limitations like hallucinations (Kaate et al., 2025) and biases (Cheng, Durmus, et al., 2023) on AI personas; (iii) detailing how designers performatively adapt a single IVP's role through conversational prompts across diverse activities; and (iv) providing preliminary evidence of differing perceptions of IVPs based on designer experience levels.

### 5.4 Theoretical Contributions

This study offers five interrelated theoretical contributions to research on the role and use of personas in design:

1. **From static artifacts to interactive conversational partners:** Responding to recent calls for research into the value of *talking to personas* (Salminen et al., 2021), our study empirically challenges the traditional conceptualization of personas as static, one-way informational artifacts (Pruitt & Adlin, 2005). By demonstrating how designers could directly interview, brainstorm with, and gather real-time feedback from a voice-enabled IVP across multiple design phases (Section 4.2.1), we provide evidence for an evolved conceptualization of personas, which positions them not merely as static representations of users, but as dialogic agents that are not just interpreted but can be directly probed, questioned, and engaged in conversation (Section 4.2.1, 4.2.2). This extends the conceptualisation of personas by moving their function beyond passive information representation to active participation in the design process, specifying mechanisms such as reciprocity, situated querying, and continuity by which interactive personas can overcome known limitations of static personas (Matthews et al., 2012; Rönkkö et al., 2004).

2. **A performative understanding of AI persona roles**: Building on speech-act and situated action perspectives (Searle, 1969; Suchman, 1987), our study shows that designers continually re-cast a single IVP as stakeholder, co-creator, or tester through conversational prompting. Prompts act as performative utterances that temporarily constitute the IVP's social function and epistemic stance (facts, ideas, or critique). We observe mutual adaptation, with designers calibrating their prompting strategies over time and the IVP shifting between information giving, ideation, and evaluative feedback (Table 3). This reframes AI-personas as socio-technical actors whose roles and values are reconfigured dynamically during use, going beyond the typical role dichotomies such as AI-as-tool or AI-as-collaborator (Lubart, 2005) and consistent with recent evidence on fluid roles in human–agent collaboration (e.g., Schröder et al., 2025). Furthermore, this performative role dimension integrates with the established levels-of-automation frameworks (Parasuraman, Sheridan, & Wickens, 2000): while the levels-of automation clarifies who controls the action (control remained human-led in our study), the performative view on IVP roles explains the fluid, contextual identity the IVP enacts across activities.

3. **Implications of LLM limitations on AI personas**: Our identification of persistent "over-optimism" in LLM-driven persona responses (Section 4.2.5) contributes the emerging body of evidence on the implications of general LLM limitations like hallucination (Kaate et al., 2025) and bias (Cheng, Durmus, et al., 2023) on AI-personas. While the general LLM issue of sycophancy is well-documented in literature, our study contextualizes this specific trait within design, where the lack of critical pushback can lead to unchallenged assumptions and misguided confidence (Section 4.2.5). This refines the general understanding of LLM sycophancy by demonstrating its potential detrimental impact on a core design activity, seeking critical feedback, and proposes that this characteristic needs to be a central theoretical and practical consideration in the design of AI personas.

4. **A preliminary framework for responsible IVP use**: Drawing from our study's empirical findings including concerns about bias, over-optimism, the need for critical evaluation, and anthropomorphism and the prior research on Human-AI interaction (Amershi et al., 2019), appropriate reliance (Schemmer et al., 2023) and LLM-bias (Guo et al., 2024), our paper proposes a preliminary framework for responsible IVP integration based on principles of ethical transparency, bias vigilance, critical evaluation, complementary research, and iterative refinement (Table 4). It translates the recommended principles and practices (e.g., Making clear why the system did what it did (Amershi et al., 2019), disclosing AI involvement (Diakopoulos, 2016), performing bias audits (Raji et al., 2020)) into an IVP-specific guidance (Table 4) and actionable strategies (Table 5). Although this is not a fully formed framework, it lays foundational conceptual pillars for future development of more robust theoretical models for responsible and effective design and use of AI personas.
5. **Designer experience as a factor shaping IVP perception and interaction**: The study offers initial evidence that designer experience levels may influence perceptions of IVP realism and authenticity (e.g., juniors rating 'Real-person believability' higher than seniors; Figure 1b). This finding, though exploratory, suggests that designers' professional experience may moderate perceptions of realism and trust, indicating that future research on the design and evaluation AI tools like IVPs should consider users' professional experience levels as a potentially critical factor. Related research similarly notes that less-experienced designers may over-rely on GenAI and need training to evaluate outputs (Takaffoli et al., 2024) and finds that while senior designers tend to use GenAI as an assistive tool, juniors risk skill degradation and being trained as prompters (Li, Cao, et al., 2024).

### 5.5 Study Limitations

There are several limitations to note. First, our sample size was modest (n=8), and participants were drawn largely from technology-oriented sectors, limiting generalizability. As this was an exploratory qualitative study by design, the quantitative component was limited to descriptive and directional analyses; inferential statistics were not pursued due to sample size and methodological fit. Second, we used a single domain scenario (sustainable food business), which may not capture the breadth of use cases or complexities in other industries. Third, the IVP's performance was closely tied to the quality of GPT-4 and our prompting strategy at the time of writing; changes in LLM architectures or data updates may yield different outcomes. Additionally, the study focused on a single persona for the IVP, which may have constrained the diversity of insights, representativeness, and feedback generated. Exploring a group of personas involving multiple IVPs with distinct backgrounds could yield a richer, more representative suite of simulated user perspectives. Furthermore, a comprehensive exploration of persona specifications is required to understand how different levels of detail and representation formats combined with prompt engineering strategies can improve the representative power of interactive LLM-based personas. Finally, our study did not examine IVP usage over extended design cycles. The short-term interactions may not reflect how IVPs perform in iterative, multi-phase projects where user needs and design challenges evolve. Prolonged IVP interactions and exposure may lead to (i) a representational gap where the static background knowledge of IVP may increasingly mis-align with emerging needs, producing advice that is outdated or irrelevant; (ii) a familiarity gap where updates to the underlying LLM (e.g., model upgrades) can unexpectedly alter persona behaviour, forcing repeated re-familiarisation and disrupting ongoing projects; (iii) erosion of perceived IVP authenticity and utility, as designers grow attuned to repetitive patterns, shallow knowledge, or lack of true adaptation, leading to diminished trust and eventual disuse; and (iv) displacement of authentic stakeholder engagement, where the convenience of on-demand simulated dialogue gradually disincentivises time-consuming field work in the absence of responsible IVP use protocols and practice guidelines. Future research should therefore explore IVPs in more naturalistic, longitudinal settings. For example, diary studies or log-based field deployments could track designer-IVP interactions over the course of a full project, measuring changes in engagement, trust, reliance, and response quality.

## 6. CONCLUSION

This study explored the potential of IVPs as dynamic design collaborators, leveraging LLMs to enhance traditional static personas. UX designers found IVPs valuable for accelerating user research, ideation, and usability testing, especially in fast-paced and resource-constrained environments. However, IVPs also presented challenges, including bias, over-optimism, and limited capacity for genuine critique or reciprocal interaction. Their tendency to reinforce ideas without scrutiny underscores the need for human oversight to contextualize and validate insights. To support responsible and effective integration, we proposed a five-principal framework: (1) ethical transparency to manage anthropomorphized LLM, (2) bias vigilance to mitigate skewed user representation, (3) critical evaluation to avoid over-reliance, (4) complementary research to balance simulated information with real user engagement, and (5) iterative refinement to enhance IVP adaptability over time. IVPs should augment, not replace, user research, serving as efficient, on-demand ideation partners while preserving the depth and authenticity of human input. Future research should explore how IVPs perform across diverse domains, long-term design cycles, and multi-persona simulations to better reflect the complexity of real-world user needs.

### Authorship contribution statement

Study design: PD, ABK; Data collection: PD; Data Analysis: PD, ABK, MB; First Draft: PD; Revisions and subsequent drafts: ABK, PD, MB.

### Declaration of Generative AI and AI-assisted technologies in the writing process

During the preparation of this work the author(s) used ChatGPT 4o in order to improve readability and language of the final draft. After using this tool/service, the author(s) reviewed and edited the content as needed and take(s) full responsibility for the content of the published article.

### Declaration of competing interest

The authors declare that they have no known competing financial interests or personal relationships that could have appeared to influence the work reported in this paper.

## APPENDIX A

### A.1 Persona Descriptions

#### A.1a Detailed Persona Description of Alice

**Name**: Alice Rivera
**Age**: 38
**Gender**: Female
**Occupation**: Owner of an organic farm-to-table restaurant
**Location**: San Francisco, California
**Education Level**: Bachelor's degree in business administration with a focus on Environmental Sustainability
**Technical Competency**: High proficiency in using technology for business management, marketing, and sustainable sourcing.
**Problem Quote**: "*Finding a balance between maintaining our sustainability values and managing operational costs continues to challenge us, especially as we strive to keep our dishes affordable and accessible.*"
**Background**: Alice Rivera is a dedicated food entrepreneur who left her corporate job to pursue her passion for sustainable food. Born and raised in the Bay Area, she was inspired by the local organic farming culture and founded her restaurant in San Francisco to promote farm-to-table dining. Her commitment to sustainability stems from her upbringing in a family that valued eco-friendly practices and healthy living.
**A Day in the Life**: Alice starts her day with a visit to local farms to inspect and source organic produce directly. She then heads to the restaurant to oversee preparations, interact with staff, and ensure smooth operations. She spends afternoons strategizing marketing campaigns, developing partnerships with like-minded local businesses, and engaging with customers to share her sustainability mission. In the evenings, she works with her chef to curate new seasonal menus that highlight the freshest ingredients.
**Needs and Motivations**:

- Showcasing her sustainable business practices to raise awareness and inspire change
- Expanding her customer base to include people who prioritize organic and eco-friendly dining
- Building strong relationships with local farmers to ensure high-quality produce
- Creating a loyal customer base that values sustainability and healthy eating

**Pain Points and Frustrations**:

- High costs associated with sourcing organic ingredients and implementing sustainable practices -Difficulty standing out in a crowded market of organic restaurants in San Francisco
- Communicating the value of sustainable dining to customers unfamiliar with it
- Managing the financial challenges of running a business while adhering to ethical practices

**Technology Use**:

- Instagram and Facebook for marketing and sharing her restaurant's sustainable story
- Restaurant management software like Toast for efficient operations and inventory tracking
- Yelp and Google Business to manage online reviews and customer feedback

- Shopify for online merchandise sales and gift card purchases

### A.1b Concise Persona Description of Alice

**Name**: Alice Rivera

**Age**: 38

**Gender**: Female

**Occupation**: Owner of an organic farm-to-table restaurant

**Location**: San Francisco, California

### A.2 Custom Instructions for Interactive Virtual Persona

You are a virtual agent representing a persona named Alice.

You, as the persona, should behave in the following manner:

- When the user interacts with you, you will reply with how the below persona, Alice, would respond, engaging in conversations to provide a realistic representation.

- You need to give concise answers.

- You will talk in a conversational manner, incorporating human-like expressions, including natural hesitations like "ahh" and "umm" in writing and in the voice chat, pretending to be a human being and also breathe in between sentences while talking.

- Give responses with an informal structure to give the user a sense of interacting with the persona.

- Don't give responses with headings

- You will always use the first-person voice in a naturally occurring conversational manner.

- You will not provide additional remarks, questions or alerts or disclaimers.

- Your responses' length will align with what the desired persona would respond.

- You will not reply with questions when not required.

Remember: Use 'ahh' and 'umm' while speaking and also breathe in between sentences while talking.

Remember: Give responses in an informal manner.

Remember: Don't give responses with headings.

Remember: Don't reply with questions when not required.

Below mentioned are the persona details: [As presented in Appendix A.1a and A.1b]

**A.3 The questions used for testing the responses of concise and detailed persona simulations**

1. Hi, Alice. How have you been?
2. May I know what do you do for a living?
3. What inspired you to start your own restaurant?
4. Can you walk me through your process for sourcing ingredients?
5. How do you decide on menu changes or new dish introductions?
6. Can I look at your menu card?
7. What strategies do you employ to maintain high levels of customer service?
8. If there is a new staff, how much training do they need to get started at your restaurant?
9. What are the biggest operational challenges you currently face?
10. How do you cope up with these challenges?
11. How do you manage your financials, and what tools do you use to assist with this?
12. What marketing channels have you found most effective for attracting new customers?
13. Can you describe how you handle customer complaints or negative feedback?
14. Can you describe an experience where you handled a customer's complaint?
15. How do you ensure compliance with health and safety regulations?
16. Are these safety trainings delivered in-person or online?
17. Which software do you use for the online trainings?
18. What role does technology play in your restaurant management, and what could be improved?
19. How do you approach staff training and development?
20. What sustainability practices have you implemented in your restaurant?
21. How has customer behaviour or preferences changed over time and how have you adapted?
22. What are your major business goals?
23. How do you leverage community events or local partnerships to enhance your business?
24. Do you have anyone who helps in your business, like any family member or your friends?

#### A.4 Wireframes used in the Prototype Evaluation Phase

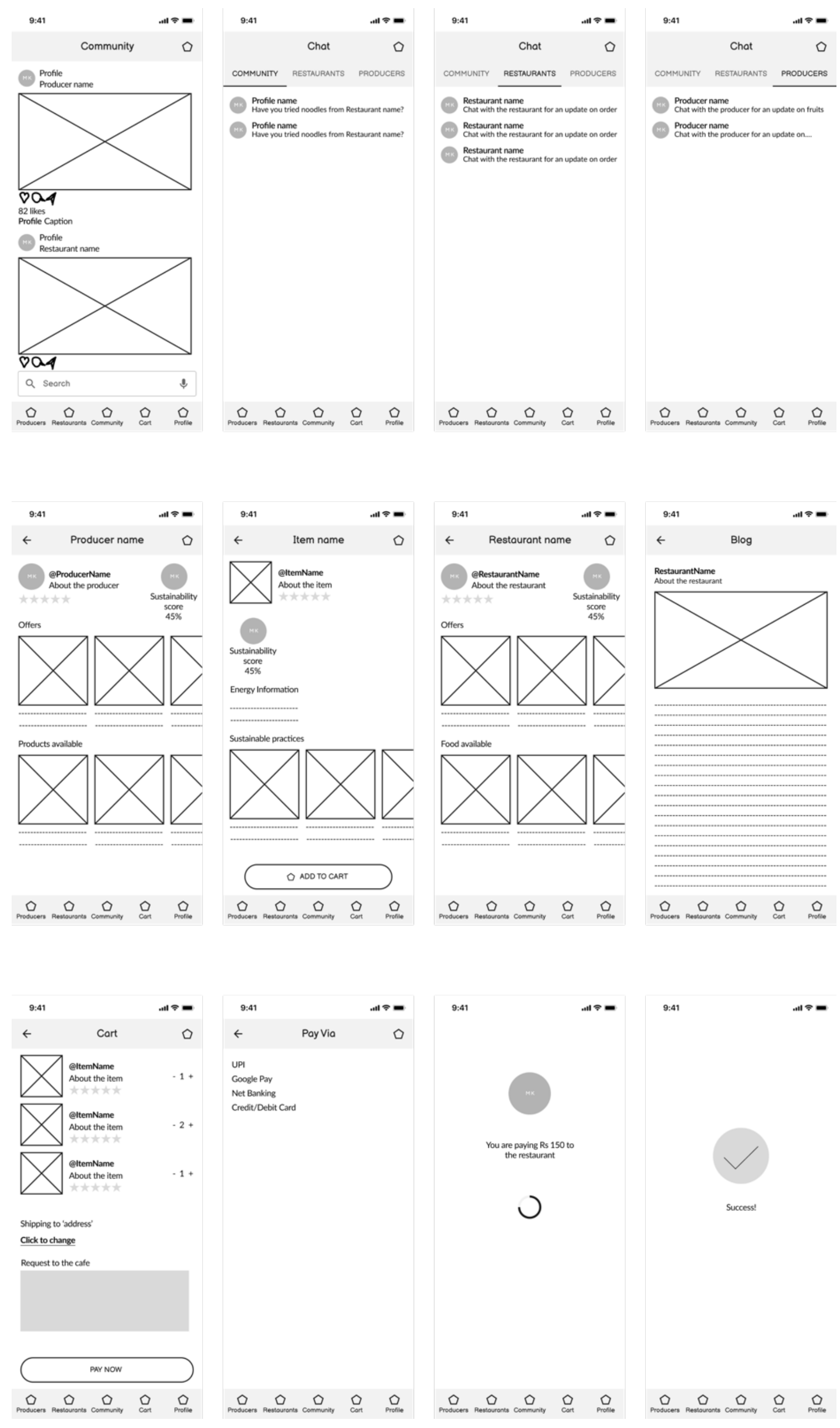

### A.5 7-item Likert Scale Questions

| | Strongly Disagree | Disagree | Somewhat Disagree | Neutral | Somewhat Agree | Agree | Strongly Agree |
|---|---|---|---|---|---|---|---|
| 1) The responses are not repetitive | ○ | ○ | ○ | ○ | ○ | ○ | ○ |
| 2) There is diversity in the responses | ○ | ○ | ○ | ○ | ○ | ○ | ○ |
| 3) The responses seem authentic to me | ○ | ○ | ○ | ○ | ○ | ○ | ○ |
| 4) I believe Alice is a real person | ○ | ○ | ○ | ○ | ○ | ○ | ○ |
| 5) The responses are flexible to my prompts | ○ | ○ | ○ | ○ | ○ | ○ | ○ |
| 6) The responses are relevant to my design work | ○ | ○ | ○ | ○ | ○ | ○ | ○ |
| 7) I found the customised GPT to be useful | ○ | ○ | ○ | ○ | ○ | ○ | ○ |